\documentclass[prl,twocolumn,superscriptaddress,showpacs]{revtex4}

\usepackage{amsmath}
\usepackage{amstext}
\usepackage{latexsym}
\usepackage[dvips]{graphicx}

\newcommand{\beq}{\begin{equation}}
\newcommand{\eeq}{\end{equation}}
\newcommand{\barr}{\begin{eqnarray}}
\newcommand{\earr}{\end{eqnarray}}
\newcommand{\ket}[1]{\left\vert#1\right\rangle}

\newcommand{\Ham}{\mathcal H}

\begin{document}

\title{Decoherence by engineered quantum baths}

\author{Davide Rossini}
\affiliation{NEST-CNR-INFM \& Scuola Normale Superiore, piazza dei
    Cavalieri 7, I-56126 Pisa, Italy}
\author{Tommaso Calarco}
\affiliation{Dipartimento di Fisica, Universit\`a di Trento and BEC-CNR-INFM,
    I-38050 Povo, Italy}
\affiliation{ITAMP, Harvard-Smithsonian Center for Astrophysics,
    Cambridge, MA 02138, USA}
\author{Vittorio Giovannetti}
\affiliation{NEST-CNR-INFM \& Scuola Normale Superiore, piazza dei
    Cavalieri 7, I-56126 Pisa, Italy}
\author{Simone Montangero}
\affiliation{NEST-CNR-INFM \& Scuola Normale Superiore, piazza dei
    Cavalieri 7, I-56126 Pisa, Italy}
\affiliation{Institut f\"ur Theoretische Festk\"orperphysik, Universit\"at Karlsruhe,
    76128 Germany}
\author{Rosario Fazio}
\affiliation{International School for Advanced Studies (SISSA)
    via  Beirut 2-4,  I-34014, Trieste - Italy}
\affiliation{NEST-CNR-INFM \& Scuola Normale Superiore, piazza dei
    Cavalieri 7, I-56126 Pisa, Italy}

\date{\today}

\begin{abstract}

We introduce, and determine decoherence for, a wide class of
non-trivial quantum spin baths which embrace Ising, XY and Heisenberg universality classes 
coupled to a two-level system.
For the XY and Ising universality classes we provide an exact expression for the decay of the 
loss of coherence beyond the case of a central spin coupled uniformly to all the spins of the baths 
which has been discussed so far in the literature. In the case of the Heisenberg spin bath 
we study the decoherence by means of the time-dependent density matrix renormalization group.
We show how these baths can be engineered, by using atoms in optical lattices.
\end{abstract}


\maketitle

Understanding decoherence is central to the description of the crossover between
quantum and classical behaviour~\cite{zurek,zurekB} and it is a crucial issue
for the successful implementation of quantum information processing~\cite{nielsen}.
Although desirable, it is not always possible to fully characterize
the bath and therefore it is necessary to resort to ingenious
modelizations. Paradigmatic models represent the environment as a set of
harmonic oscillators~\cite{weiss} or spin-1/2 particles~\cite{prokofiev}.
In order to grasp all the subtleties of the entanglement between the system and
its environment it would be of great importance to study engineered baths (and
system-bath interaction) that can be realized experimentally and whose properties are
amenable of an exact solution. Here we discuss a new class of spin
baths which satisfy these requirements. We show how to realize them by means of optical
lattices and we find the exact solution for the decoherence.
We thus provide a way to test in the laboratory the emergence of decoherence
in quantum systems.

\begin{figure}[!h]
    \includegraphics[scale=0.5]{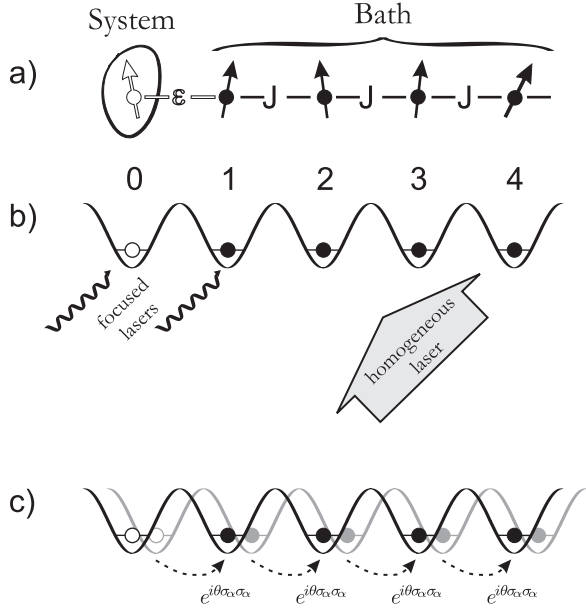}
    \caption{a) A sketch of the system plus bath model we consider in this
    work. The two level system (at position zero) is coupled to the
    $\sigma_z$ component of the first spin of the chain that acts as
    a spin bath. Atoms in an optical lattice can simulate
    this controlled decoherence by means of series of lasers b) and
    displacements of the lattice c) which allow to realize both the
    interaction of the bath with an external magnetic field and the
    anisotropic exchange coupling present in Eq.~\eqref{Hbath}.}
    \label{fig:scheme}
\end{figure}

The setup we consider is shown in Fig.~\ref{fig:scheme}. Here a two-level system
(the quantum system) is coupled to a single spin of a one-dimensional spin-1/2 chain
(the environment). The free evolution of the chain is described by means of a
Hamiltonian which embraces Ising, XY and Heisenberg universality classes and it is
therefore well understood. 

In this context we analyze how decoherence induced on the two-level system depends on the 
internal environment dynamics. In particular by varying the parameters of the optical lattice 
we can test what happens when the bath enters different phases (critical, ferromagnetic,
anti-ferromagnetic, etc$\dots$). The motivation to analyze this model stems from
the interest in the decoherence due to spin baths both in the absence~\cite{zurek2,cucchietti}
and in the presence~\cite{tessieri,dawson,paganelli,quan,cucchietti2,khveshchenko}
of interaction among the spins of the bath. In particular
Dawson {\em et al.} suggested that, due to the monogamy of entanglement,
decoherence depends strongly on the entanglement present within the bath~\cite{dawson}.
The model presented here allows us to frame this observation in a broader
context, by pointing out different aspects between the internal entanglement
of the bath and the induced decoherence of the quantum system. We believe that
our analysis may provide a fairly general understanding of the relation
between decoherence and ground state properties of spin baths.
We will first present the exact solution of the theoretical problem and then
discuss its experimental implementation.

{\underline {\em The Model} } -
The system-bath model shown in Fig.~\ref{fig:scheme} is described by the
Hamiltonian
$
    \Ham = \Ham_{TL} + \Ham_{E} +  \Ham_{IN} \,\,.
$
%
%
%
The two-level system, characterized by the ground state $\ket{g}$ and the excited
state $\ket{e}$, has  a free Hamiltonian ${\cal H}_{TL}=\omega_e |e\rangle\langle e|$
and is coupled to the first spin of the bath through the interaction
${\cal  H}_{IN} = - \epsilon |e\rangle \langle e| \sigma_1^z$
with $\epsilon$ being the coupling constant. 
The Hamiltonian of the environment is
\begin{eqnarray}
    \Ham_{E} &= &- \frac{J}{2} \sum_{j=1}^N \left[
    \left(  1 + \gamma \right) \sigma^x_j \sigma^x_{j+1} +
    \left( 1 - \gamma \right) \sigma^y_j \sigma^y_{j+1} + \right.
    \nonumber \\
    & &
    \left.
    \Delta \sigma^z_j \sigma^z_{j+1}
    + 2 \lambda \sigma^z_j \right]
\label{Hbath}
\end{eqnarray}
where $\sigma^{\alpha}_i$ ($\alpha = x,y,z$) are the Pauli matrices of the $i$-th spin.
The constants $J$, $\Delta$ and $\lambda$ represent the exchange coupling,
the anisotropy parameter and an external  magnetic field. The model defined by
Eq.~\eqref{Hbath} has a very rich structure~\cite{qptbook}.
We consider the two separate cases of $\{ \Delta =0, \lambda \ne 0 \}$ and
$\{ \Delta  \ne 0 , \lambda = 0 \}$.
When the anisotropy parameter is set to zero,
and $0 < \gamma \leq 1$, the model of Eq.~\eqref{Hbath} belongs to the Ising universality 
class which has a critical point at $\lambda_c = 1$; when $\gamma = 0$ it belongs to the XY 
universality class. 
At $\lambda =0$, the model of Eq.~\eqref{Hbath} is critical if $-1 \leq \Delta \leq 1$, in the other
cases the chain has ferromagnetic or anti-ferromagnetic order if the anisotropy is
positive or negative respectively. Our model goes beyond the model of a central spin coupled 
uniformly with all the spins of the bath and can be more easily simulated with cold atoms as 
discussed towards the end of this paper.

The evolution of the reduced density matrix $\rho$ of the two-level system
corresponds to a purely dephasing process.
In the basis of the eigenstates $\ket{g}, \, \ket{e}$, the diagonal terms
$\rho_{gg}(t)$ and $\rho_{ee}(t)$ do not evolve in time.
Only the off-diagonal terms will decay according to the expression
$ \rho_{eg} (t) = \rho_{eg} (0) D(t)$ where 
$D(t)= \langle \vert e^{i \Ham t} e^{-i (\Ham_{TL} + \Ham_{E}) t} \vert \rangle$ is 
evaluated over the ground state of the spin bath.
The decoherence of the system is fully captured by the function ${\cal L}(t)=| D(t) |^2$
sometimes called Loschmidt echo.
In the following we will study the dependence of ${\cal L}(t)$ upon the spin bath
Hamiltonian both in the case of periodic and open boundary conditions
for a number $N$ of bath spins of the order of $\sim 10^2 -10^3$.

In the case $\Delta =0$ and $\lambda \ne 0$, the function ${\cal L}(t)$ can be calculated
exactly.
By means of the Jordan-Wigner transformation $\sigma^+_j = c^\dagger_j \exp{(i \pi
\sum_{k=1}^{j-1}c^\dagger_k c_k)}$, and  $\sigma^z_j = 2c^\dagger_jc_j -1$, it is
possible to map the Hamiltonian of the spin bath onto a free fermion model
which can be expressed in the form  $\Ham_{E} = \frac{1}{2} \mathbf{\Psi^\dagger C \Psi}$, 
where $\mathbf{\Psi^\dagger} = ( c^\dagger_1 \ldots c^\dagger_N \, c_1 \ldots c_N)$
($c_i$ are the corresponding spinless fermion
operators) and $\mathbf{C} = \sigma^z \otimes \mathbf{A} + i \sigma^y \otimes \mathbf{B}$
is a tridiagonal block matrix with
$A_{j,k} = -J (\delta_{k,j+1} + \delta_{j,k+1} ) -2 (\lambda + \epsilon_j )
\delta_{j,k}$ and $B_{j,k}= -J \gamma \, ( \delta_{k,j+1} - \delta_{j,k+1})$.
The Loschmidt echo can then be evaluated~\cite{levitov,klich} exactly:
\beq
    {\cal L}(t) =
    \mathrm{det} \left( 1 - \mathbf{r} + \mathbf{r} e^{i \mathbf{C} t} \right) 
\label{Lising}
\eeq
where $\mathbf{r}$ is a matrix whose elements 
$r_{i,j} = \langle \Psi_i^\dagger \Psi_j \rangle$
are the two-point correlation functions of the spin chain.
Equation (\ref{Lising}) allows us to go beyond the central spin model considered so far in
the literature and enables us to address explicitly the case of a large number of
spins in the bath.

\begin{figure}[!t]
   \includegraphics[scale=0.4]{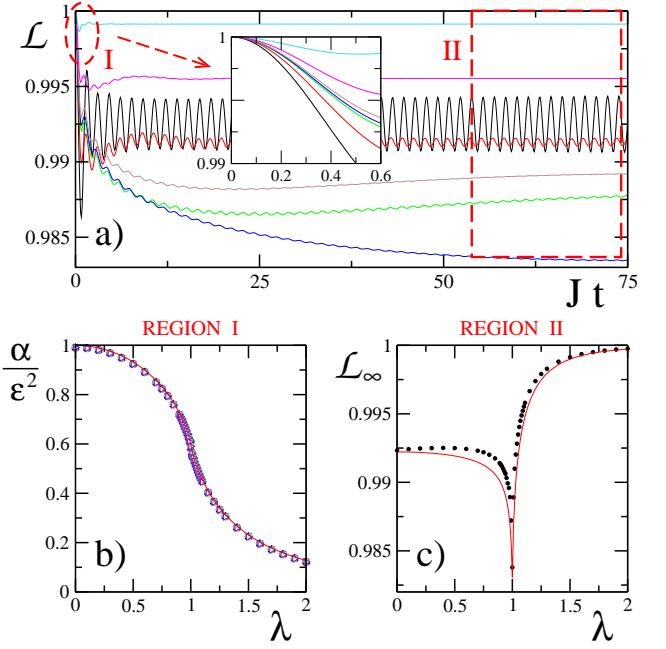}
    \caption{a) Loschmidt echo as a function of time for a qubit coupled
        to a $N=300$ spin Ising chain; $\epsilon=0.25$.
        The various curves are for different values of the transverse
        magnetic field: $\lambda=$ 0.5 (black), 0.9 (red), 0.99 (green),
        1 (blue), 1.01 (brown), 1.1 (magenta), 1.5 (cyan).
        Inset: zoom in at small times.
        b) In the region I, ${\cal L}$ has a Gaussian decay with a typical
        scale $\alpha$. Here we plot $ \alpha / \epsilon^2$
        as a function of $\lambda$ for different values of $\epsilon$:
	0.01 (circles), 0.025 (squares), 0.05 (diamonds),
        0.075 (triangles up), 0.1 (triangles left), 0.25 (triangles down),
        0.5 (triangles right).
	The solid line shows the result of a perturbative calculation
	at small times.
    	c) In the region II, ${\cal L}$ oscillates around a constant value 
	${\cal L}_{\infty}$. Here we plot this constant background as a 
	function of $\lambda$ for
    	$\epsilon=0.25$. Points indicate the data from the exact solution, while
	the solid line is the result of the perturbation expansion
	in the coupling constant $\epsilon$.}
    \label{fig:LEcho_N300}
\end{figure}

Fig.~\ref{fig:LEcho_N300}a shows the generic behaviour of ${\cal L}$ as
a function of time for different values of $\lambda$ ($\gamma=1$).
For $\lambda < 1$ the echo oscillates with a frequency proportional
to $\epsilon$, while for $\lambda > 1$ the amplitudes of oscillations
are drastically reduced. As the chain is finite, at long times there are
revivals, but for $N \sim 10^3$ there is already a wide interval
(region II) where the asymptotic behaviour can be analyzed.
%
%
%
%
It is useful to consider in detail the short- and long-time behaviour (regions
I and II in Fig.~\ref{fig:LEcho_N300}a) of ${\cal L}$.
At small times the decay is Gaussian, ${\cal L}_I(t) \sim  e^{-\alpha t^2}$
(region I is expanded in the inset of Fig.~\ref{fig:LEcho_N300}a).
The scale of the Gaussian decay at short times displays a remarkable universal
behaviour shown in Fig.~\ref{fig:LEcho_N300}b where
$\alpha (\lambda, \epsilon)$ is shown as a function of $\lambda $.
At long times (Region II) and for $\lambda > 1 $ the Loschmidt echo approaches
an asymptotic value ${\cal L}_{\infty}$, while for $\lambda < 1$ it oscillates
around a value which is constant in time (see Fig.~\ref{fig:LEcho_N300}a for a
qualitative picture). This limiting value ${\cal L}_{\infty}$ strongly depends
on $\lambda$ as shown in Fig.~\ref{fig:LEcho_N300}c. Evidence that
${\cal L}_{\infty}$ describes the asymptotic regime can be obtained by comparing
the data with the result of an analytical expression based on a perturbation
expansion in the coupling,
$
{\cal L}_{\infty} \sim [ 1 - (\epsilon^2/2) \sum_{k \neq 0} | \langle \psi_k
\vert \,  \sigma^z_1\, \vert \psi_0 \rangle |^2
/( E_k - E_0 )^{2}]^2
$
where $\vert \psi_{k/0} \rangle$ are the excited states (ground state) of $\Ham_E$
with energy $E_{k/0}$.
Results obtained from the previous expression are plotted in
Fig.~\ref{fig:LEcho_N300}c (solid line) together with the exact solution (points).

\begin{figure}[!h]
  \begin{center}
    \includegraphics[scale=0.35]{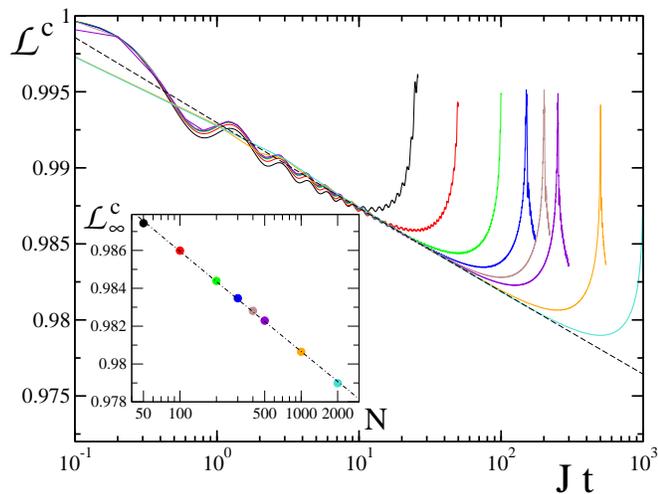}
    \caption{Loschmidt echo as a function of time at the critical point
        $\lambda = \lambda_c$, for different sizes of the chain: $N=50$ (black),
        100 (red), 200 (green), 300 (blue), 400 (brown), 500 (violet),
        1000 (orange), 2000 (cyan) and at
        $\epsilon = 0.25$. The dashed line is a fit with a logarithmic decay of the type
	${\cal L}^c(t) = c_0 / (1+ c_1 \, \textrm{ln} \, t)$.
	Inset: Minimum value of ${\cal L}^c_{\infty}$ as a function of $N$.
	Numerical data have been fitted with Eq.~\eqref{asimp} (dotted-dashed line),
        where $l_{\infty} \approx 0.997, \; \beta \approx 2.369 \times 10^{-3}$.}
    \label{fig:L0critic}
  \end{center}
\end{figure}

Additional information emerge when analyzing the scaling of ${\cal L}$
with the size $N$ of the chain. If the chain is not at the critical point
and $N$ is large, both the $\alpha$ and the saturation value ${\cal L}_{\infty}$ are
almost independent of the bath size.
At the critical point the situation is rather peculiar (In Fig.~\ref{fig:L0critic}
we plot the decay of ${\cal L}$ at $\lambda_c$, for different sizes of the chain).
The decay is very slow in time ${\cal L}^c(t)  \sim \textrm{ln}^{-1} \, t$,
moreover the minimum value ${\cal L}_{\infty}^c$ reached by the Loschmidt echo depends
on $N$ as
\beq
{\cal L}_{\infty}^c = \frac{l_{\infty}}{1+ \beta \ln N} \, .
\label{asimp}
\eeq
All the properties described so far are typical as long as
$0 < \gamma \leq 1$ (the model belongs to the Ising universality
class).
If the chain is described by  the $XY$ model ($\gamma = 0$)
the Loschmidt echo behaves as in Eq.~\eqref{asimp} for $\lambda < 1$
(the model is critical in this range).  For $\lambda > 1$ we found ${\cal L} (t) = 1$;
in this case the coupled qubit does not decohere at all.

\begin{figure}[!h]
    \includegraphics[scale=0.31]{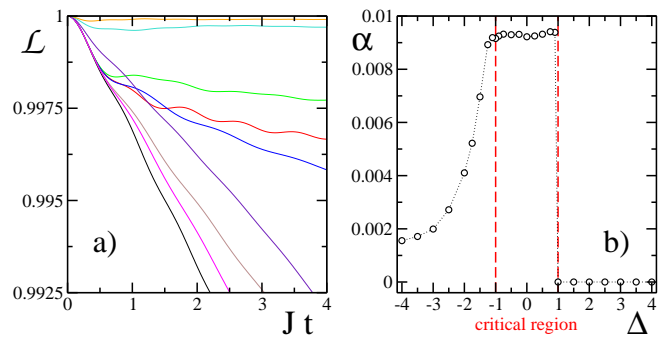}
    \caption{a) Loschmidt echo for a qubit coupled to
        an $N=100$ spin XXZ Heisenberg chain; $\epsilon = 0.1$ and
        $\Delta =$ 0.9 (black), 0.5 (red), 0 (green), -0.5 (blue),
        -0.9 (brown), -1 (magenta), -1.5 (violet), -2 (cyan), -2.5 (orange).
	b) Decay rate $\alpha$ of the Loschmidt echo at small times
        as a function of $\Delta$; the dashed lines indicate the critical
        region $-1 \leq \Delta \leq 1$.}
    \label{fig:HEIS_fid}
\end{figure}

In the other case we consider ($\Delta \ne 0$, $\lambda = 0$)
an analytical solution is not available. We then resort to
the recently developed time dependent Density Matrix
Renormalization Group (t-DMRG)~\cite{vidal2,white} with open boundary
conditions.
The results shown in Fig.~\ref{fig:HEIS_fid}a have been obtained
for chains of $N=100$ spins and different values of the anisotropy $\Delta$.
We have set a Trotter slicing $dt=10^{-2}$ and a truncated Hilbert space
of dimensions $m=100$.
Also in this case the Loschmidt echo is Gaussian at short times and the scale $\alpha$ of
this decay  for the Heisenberg model is shown in Fig.~\ref{fig:HEIS_fid}b.
The dependence of $\alpha$ on the different regions of the phase diagram is
evident. The rate of decoherence is strong in the critical region, $-1 \le \Delta \le 1$,
because of the presence of low energy modes.
In the ferromagnetic phase, $\Delta \ge 1$, ${\cal L}$ does not decay because the
bath is fully polarized.
%
%
In Fig.~\ref{fig:HEIS_fid}b we plot the decay rate $\alpha$ as a function of $\Delta$.
The dependence on the critical properties of the chain is again remarkable.
We notice two qualitatively different behaviours at the boundaries
of the critical region: at $\Delta=+1$ there is a sharp discontinuity,
while at $\Delta=-1$ the curve is continuous; in the critical region
$\alpha$ is constant.

The rich structure of entanglement in the ground state of Hamiltonian (\ref{Hbath}) 
suggests the following picture. When the decay is quadratic, $Jt \ll 1$, only 
short-range correlations in the bath are important and it is therefore natural 
to relate $\alpha$ to the nearest-neighbour concurrence~\cite{osborne,osterloh}. 
We checked this both for the Ising and the XXZ model and we found that $\alpha$ is 
proportional to the concurrence for $\lambda \ge 1$ for the Ising case and 
$\Delta \le -1$ for the XXZ case. In the long-time limit long-range correlations 
become important, in this case the entanglement measured by the block 
entropy~\cite{vidal} or the localizable entanglement~\cite{verstraete} seems to 
be relevant to the decoherence process.

\noindent {\underline {\em Experimental implementations } } - The
properties of decoherence discussed so far can be tested
experimentally in optical lattices. Following the idea of a
Universal Quantum Simulator described in~\cite{UQS}, the time
evolution operator associated with ${\cal H}$ over a time $t$ can
be simulated by decomposing the evolution into a product of
operators acting on very short times $\tau\ll t$. In our case the
operations required are $U^z_j(\theta)\equiv
e^{i\theta\sigma_j^z}$, and $U^{\alpha\beta}_{jk}(\theta)\equiv
e^{i\theta\sigma_j^\alpha\sigma_k^\beta}$. For $\alpha\in\{x,y\}$
one can write $U^{\alpha\alpha}_{jk}=V^\alpha_jV^\alpha_k
U^{zz}_{jk}V^{\alpha\dagger}_kV^{\alpha\dagger}_j$. Here,
$
V^\alpha_j=(\openone-i\sigma^\alpha_j)/\sqrt{2}
$
are fast homogeneous local unitary operations that can be
realized with single atoms trapped in an optical
lattice~\cite{UQS}, each having two relevant electronic levels
($|0\rangle_j$,
$|1\rangle_{j}$) interacting with a resonant laser. These
operations can be made very fast by simply increasing the laser
intensity and thereby the Rabi frequency. They can be performed
either simultaneously on all qubits, by shining the laser
homogeneously onto all atoms, or selectively on some of them, by
focusing it appropriately (see Fig.~\ref{fig:scheme}b). For our
purposes the individual addressing is needed only for the atom in
position $0$, which represents the quantum system. Two-qubit
operations can be performed by displacing the lattice in a
state-selective way~\cite{Jaksch}, so that state
$|0\rangle_j|1\rangle_{j+1}$ acquires a phase factor
$e^{-i\varphi}$, as experimentally realized in~\cite{Bloch}. The resulting gate
$G_{j,j+1}(\varphi)$ can be composed with $\sigma^x$ rotations to yield
$U^{zz}_{j,j+1}(\theta)=e^{i\theta}[G_{j,j+1}(2\theta)\sigma^x_j\sigma^x_{j+1}]^2$.
Since we want a different coupling for the $\{01\}$ pair than for all others, we
need to erase the effect of the interaction for that specific pair
using only local operations on atom $0$, as in the sequences
$
[\sigma^z_0U^{xx}_{01}(\theta)]^2=[\sigma^z_0U^{yy}_{01}(\theta)]^2=
[\sigma^x_0U^{zz}_{01}(\theta)]^2=\openone.
$
To generate each simulation step, we shall first apply the
$-\epsilon\sigma_z\sigma_z$ term in Eq.~(\ref{Hbath}) on all atoms
(plus possibly other single-qubit terms to simulate the transverse
field) and then the remainder of two-qubit terms --~including
$(\epsilon-J\Delta/2)\sigma_z\sigma_z$~-- while taking care of washing
out the phases in the $(01)$ pair as described above.
An alternative scheme~\cite{Duan}, based on tunnel coupling between neighbouring
atoms rather than on lattice displacements, can attain the
simulation described here for the special case $\gamma=0$, but would require 
some additional stroboscopic steps in order to reproduce the general case.
Moreover, an additional optical
tweezer holding atom $0$ would be needed in this scheme to achieve
a system-bath coupling different from the intra-bath one.

Finally we would like to comment on possible extensions of this
approach. It would be quite interesting to consider as an
engineered bath a three-dimensional optical lattice. Beside being
feasible from an experimental point of view, this could be useful
in studying for instance the situation found in solid-state NMR~\cite{cory}.
It would be also intriguing to study the Bose-Hubbard model as a
bath, which would make the experimental realization even simpler.
Here we chose to focus on spin baths, as they are central to the
study of decoherence in many quantum systems.

We want to thank D. Cory, G. De Chiara and M. Rizzi for very
useful discussions. We acknowledge support from EC (grants
RTNNANO, SPINTRONICS, ACQP, SCALA and EUROSQIP), MIUR-PRIN, 
NSF (grant of the ITAMP at Harvard University and Smithsonian 
Astrophysical Observatory) and Centro di Ricerca Matematica 
``Ennio De Giorgi'' of SNS.


\begin{thebibliography}{99}

\bibitem{zurek}
    W.H. Zurek, Phys. Today {\bf 44}, 36 (1991)
\bibitem{zurekB}
    W.H. Zurek, Rev. Mod. Phys. {\bf 75}, 715 (2003).
\bibitem{nielsen}
    M.A. Nielsen and I.L. Chuang,
    {\it Quantum Computation and Quantum Information}
    (Cambridge University Press, Cambridge, 2000).
\bibitem{weiss}
    U. Weiss, {\it Quantum dissipative systems}, 2nd edition.
    (World Scientific, Singapore, 1999).
\bibitem{prokofiev}
    N.V. Prokof'ev and P.C.E. Stamp,
    Rep. Prog. Phys. {\bf 63}, 669 (2000).
\bibitem{zurek2}
     W.H. Zurek, Phys. Rev. D {\bf 26}, 1862 (1982)
\bibitem{cucchietti}
    F.M. Cucchietti, J.P Paz, and W.H. Zurek,
    Phys. Rev. A {\bf 72}, 052113 (2005)
\bibitem{tessieri}
    L. Tessieri and J. Wilkie,
    J. Phys. A {\bf 36}, 12305 (2003).
\bibitem{dawson}
    C.M. Dawson {\it et al.},
    Phys. Rev. A {\bf 71}, 052321 (2005).
\bibitem{paganelli}
    S. Paganelli, F. de Pasquale, and S.M. Giampaolo,
    Phys. Rev. A {\bf 66}, 052317 (2002).
\bibitem{quan}
    H.T. Quan {\it et al.},
    Phys. Rev. Lett. {\bf 96}, 140604 (2006).
\bibitem{cucchietti2}
    F.M. Cucchietti, S. Fernandez-Vidal, and J.P. Paz,
    quant-ph/0604136.
\bibitem{khveshchenko}
    D.V. Khveshchenko,
    Phys. Rev. B {\bf 68}, 193307 (2003).
\bibitem{qptbook}
    S. Sachdev, {\it Quantum Phase Transitions}
    (Cambridge University Press, Cambridge, 2000).
\bibitem{levitov}
    L.S. Levitov, H. Lee, and G.B. Lesovik,
    J. Math. Phys. {\bf 37}, 4845 (1996).
\bibitem{klich}
    I. Klich,
    in {\em Quantum Noise in Mesoscopic Physics}, Nazarov Yu.V. Ed.,
    NATO Science Series, Vol. 97 (Kluwer Academic Press, 2003).
\bibitem{vidal2}
    G. Vidal, Phys. Rev. Lett. {\bf 93}, 040502 (2004)
\bibitem{white}
    S.R. White and A.E. Feiguin,
    Phys. Rev. Lett. {\bf 93}, 076401 (2004).
\bibitem{osborne}
    T.J. Osborne and M.A. Nielsen,
    Phys. Rev. A {\bf 66}, 032110 (2002).
\bibitem{osterloh}
    A. Osterloh {\it et al.},
    Nature {\bf 416}, 608-610 (2002).
\bibitem{vidal}
     G. Vidal {\it et al.},
    Phys. Rev. Lett. {\bf 90}, 227902 (2003).
\bibitem{verstraete}
    F. Verstraete, M.A. Martin-Delgado, and J.I. Cirac,
    Phys. Rev. Lett. {\bf 92}, 087201 (2004).
\bibitem{UQS}
    E. Jan\'e {\it et al.},
    Quantum Inf. and Comp. {\bf 3}, 15 (2003).
\bibitem{Jaksch}
    D. Jaksch {\it et al.},
    Phys. Rev. Lett. {\bf 82}, 1975 (1999).
\bibitem{Bloch}
    O. Mandel {\it et al.},
    Nature {\bf 425}, 937-940 (2003).
\bibitem{Duan}
    L.-M. Duan, E. Demler, and M.D. Lukin,
    Phys. Rev. Lett.  {\bf 91}, 090402 (2003).
\bibitem{cory}
    D.G. Cory {\it et al.},
    Fortschr. Phys. {\bf 48}, 875 (2000).
\end{thebibliography}
\end{document}